\documentclass[runningheads]{llncs}
\usepackage{graphicx}
\usepackage{caption, subcaption}

\begin{document}
\title{The Rise of GitHub in Scholarly Publications\thanks{Supported by the Alfred P. Sloan Foundation, \url{https://sloan.org/grant-detail/9628}}}

\author{Emily Escamilla\inst{1}\orcidID{0000-0003-3845-7842} \and
Martin Klein\inst{2}\orcidID{0000-0003-0130-2097} \and
Talya Cooper\inst{3}\orcidID{0000-0003-4241-6330} \and
Vicky Rampin\inst{3}\orcidID{0000-0003-4298-168X} \and 
Michele C. Weigle\inst{1}\orcidID{0000-0002-2787-7166} \and
Michael L. Nelson\inst{1}\orcidID{0000-0003-3749-8116}}
\authorrunning{E. Escamilla et al.}

\institute{Old Dominion University, Norfolk, VA, USA \\
\email{evogt001@odu.edu, \{mweigle, mln\}@cs.odu.edu} \and
Los Alamos National Laboratory, Los Alamos, NM, USA \\
\email{mklein@lanl.gov} \and
New York University, New York, NY, USA \\
\email{\{tc3602, vs77\}@nyu.edu}
}

\maketitle              
\begin{abstract}
The definition of scholarly content has expanded to include the data and source code that contribute to a publication. While major archiving efforts to preserve conventional scholarly content, typically in PDFs (e.g., LOCKSS, CLOCKSS, Portico), are underway, no analogous effort has yet emerged to preserve the data and code referenced in those PDFs, particularly the scholarly code hosted online on Git Hosting Platforms (GHPs). Similarly, the Software Heritage Foundation is working to archive public source code, but there is value in archiving the issue threads, pull requests, and wikis that provide important context to the code while maintaining their original URLs. In current implementations, source code and its ephemera are not preserved, which presents a problem for scholarly projects where reproducibility matters. To understand and quantify the scope of this issue, we analyzed the use of GHP URIs in the arXiv and PMC corpora from January 2007 to December 2021. In total, there were 253,590 URIs to GitHub, SourceForge, Bitbucket, and GitLab repositories across the 2.66 million publications in the corpora. We found that GitHub, GitLab, SourceForge, and Bitbucket were collectively linked to 160 times in 2007 and 76,746 times in 2021. In 2021, one out of five publications in the arXiv corpus included a URI to GitHub. The complexity of GHPs like GitHub is not amenable to conventional Web archiving techniques. Therefore, the growing use of GHPs in scholarly publications points to an urgent and growing need for dedicated efforts to archive their holdings in order to preserve research code and its scholarly ephemera.

\end{abstract}
\section{Introduction}
Researchers increasingly use and create open source software as a part of their scholarship, making software a vital element of our scholarly record. A 2014 survey by the Software Sustainability Institute found that 92\% of academic respondents use research software and 56\% developed their own software \cite{hettrick}. These researchers rely on tools such as version control systems and repository hosting platforms to develop, reuse, version, and share software. A version control system (VCS) is a tool that helps users manage changes to a repository over time. A typical code repository contains a set of files, such as program and configuration files. Web-based repository hosting platform services let users host their code projects remotely. Repository hosting platforms also provide collaborative features, including discussion threads, and allow for edits and contributions by outside collaborators.

A study by Fäber \cite{farber-jcdl2020} found that GitHub,\footnote{https://github.com} a Git Hosting Platform (GHP), was the most popular repository hosting platform. A GHP is a type of repository hosting platform made specifically for Git VCS. These include platforms such as GitHub, GitLab, and SourceForge. The increased use of Git and GHPs in academia represents a victory for open access scholarship and for computational reproducibility. We believe that when people share code openly and receive credit for it (for example, through citations), potentially leading to novel collaboration and funding endeavors, open science benefits. These platforms allow for a number of scholarly activities like peer review; however, most lack a preservation plan. The fact that some VCSs have already been discontinued -- Gitorious (2014), Google Code (2016), among others \cite{squire} -- points to the urgency of the need for a more concerted preservation effort. In addition to sustainability concerns with the platforms themselves, few workflows, tools, and processes exist for preservation of research code, as they do for other scholarly materials such as papers, data, and media \cite{reich-dlib2001,reich-serials2008,he-aslib2016,fenton-serials2006}. As the use of Git and GHPs rises amongst researchers, it becomes more important to preserve research code in order to prevent gaps in this part of the scholarly record. Often these repositories represent the bulk of a scholar's time and efforts in their research. As such, these materials are key for verifying, reproducing, and building on each others' scholarly contributions. 

In this paper, we present work that finds research software as it is represented in literature, quantifies its impact in the scholarly record, and provides a stronger basis for addressing the long-term sustainability of scholarly code and its contextual, scholarly ephemera. While source code repositories are not always included in the references of a publication, links to repositories appear throughout scholarly manuscripts as part of the evidence and support for the work being presented. We analyzed the arXiv and PubMed Central (PMC) corpora to determine the extent to which publications reference the Web at large and reference GHPs, specifically GitHub, GitLab, SourceForge, and Bitbucket. In both arXiv and PubMed Central, we found that the average number of URIs in a publication has steadily increased, as has the number of links to GHPs. In the arXiv corpus, one out of five publications contains a reference to GitHub, the most popular GHP in arXiv and PMC. 

\section{Related Work}
This is one of few studies that looks at the representation of links \textit{to} scholarly source code in scholarly literature. Previous works have investigated the opposite: representation of links \textit{to} scholarly literature \textit{from} scholarly source code repositories. Wattanakriengkrai et al. \cite{wattanakriengkrai_github_2022} studied the extent to which scholarly papers are cited in public GitHub repositories  to gain key insights into the landscape of scholarly source code production, and uncovered potential problems with long-term access, tracing, and evolution of these repositories. Färber \cite{farber-jcdl2020} analyzed data from Microsoft Academic Graph, which attempts to map publications to their source code repositories, in order to look at the content and popularity of academic source code related to published work. Färber's work focuses on the content of the GitHub repositories referenced in scholarly publications, while our work looks at how scholarly publications link to GHPs. Other related work addresses finding scholarly source code repositories hosted on GHP, either by looking through the content of the repository or by searching for links to scholarly literature in the repositories themselves. Hasselbring et al. \cite{hasselbring} investigated public repositories on GitHub and estimated that it contained over 5,000 repositories of specifically research software -- a similar estimation to Färber.

Understanding the extent to which scholarly articles reference source code is important because scholarly materials that are hosted on the Web are vulnerable to decay in the same manner as Web resources in general. In 2014, Klein et al. \cite{klein-plos2014} analyzed the use of URIs to the Web at large in the arXiv, Elsevier, and PMC corpora from 1997 to 2012. They found that the number of general URIs used in scholarly publications rapidly increased from 1997 to 2012. However, they also found that reference rot affects nearly 20\% of Science, Technology, and Medicine (STM) publications. When looking specifically at publications with a Web reference, seven out of ten publications are affected by reference rot. Reference rot is a general term that indicates that either link rot or content drift has altered the content of the Web page to be different than the content to which the author was originally referring \cite{vandesompel-icm2014}. Link rot occurs when the URI that was originally referenced is completely inaccessible. Link rot can cause the ``404: Page not found'' error that most Web users have experienced. Reference rot is caused by the dynamic and ephemeral nature of the Web. Content drift occurs when the content that was originally referenced by a URI is different from the content currently available at the URI. Jones et al. \cite{jones-plos2016} found that 75\% of references suffer from content drift. Additionally, they found that the occurrence and impact of content drift increases over time. In 2015, only 25\% of referenced resources from 2012 publications were unchanged and, worse yet, only 10\% of publications from 2006 were unchanged.

Understanding the scope of how scholarly source code is represented in scholarly literature is vital to strengthening efforts to preserve and make this code available for the long-term, as a part of the scholarly record. Some researchers attempt to make their code available for the long-term by self-archiving: depositing their own materials into a repository or archive. However, of academics who write source code, only 47.2\% self-archive that code \cite{iasge_hicss}. While self-archiving can help safeguard research software, it has not yet become part of scholars' routines.

Zenodo,\footnote{https://zenodo.org} a non-profit repository maintained by CERN that supports open data and open access to digital scholarly resources, is one example of a repository with specific functionality to support researchers who wish to self-archiving their code for long-term access. Zenodo provides a webhook that allows users to deposit new releases from GitHub repositories. Zenodo makes a copy of the code, rather than simply linking out to the GitHub page, creates relationships to previous and subsequent versions of the code, and mints a DOI for the record with software-specific metadata attached. 

Other approaches aim to ensure long-term access to scholarly code without relying on researchers doing preservation work themselves. The non-profit Software Heritage\footnote{https://softwareheritage.org/} conducts programmatic captures of public source code on the web with the goal ``to collect, preserve, and share all software that is publicly available in source code form". As part of this goal, the content of Google Code, which was phased out in 2016, is contained in Software Heritage \cite{dicosmo-ipres2017}.

However, software and code are not sufficient as stand-alone products, especially in a scholarly context where reproducibility matters. For instance, documentation about installation and dependencies are crucial for secondary users who want to reproduce and build on research. In addition, many projects maintain discussions, wikis, and other contextual items that make the source code more comprehensible and reusable for others. When referring to scholarly source code, we call these materials \textit{scholarly ephemera} \cite{iasge_enviro_scan}. Scholarly ephemera housed with a repository on a GHP (e.g., Issues on GitHub) include useful, preservation-worthy information, such as peer review, discussion of important implementation details, and questions from secondary users of the scholarly code.

Presently, neither self-archived code nor programmatically captured code incorporates the scholarly ephemera that can help secondary readers understand and evaluate the source code being cited. This is where Web archiving may be beneficial. Web archiving's goal lies in preserving the Web so that users can see a Web page as it existed at a certain point in time, which is helpful for archiving source code and the accompanying scholarly ephemera. However, because of the resources it takes to archive the Web, automated Web archiving services like the Internet Archive will crawl the most visited Web pages frequently, while the least visited Web pages, including scholarly content, may never be fully captured. Although the Internet Archive includes some GHP sites, it cannot be depended upon to preserve any given page in its entirety. Other Web archiving tools like the Webrecorder suite \cite{webrecorder} provide higher quality captures of source code and ephemera, but take more time, resulting in decreased scalability for archiving the Web at large. Also, while current Web archiving implementations are well-suited for archiving the scholarly ephemera around scholarly code, they are less effective with the source code itself, which has different metadata and reuse needs than a typical Web page. 

We know that: a) materials hosted on the Web and cited in scholarly literature are subject to reference rot, b) source code and its important scholarly ephemera are particularly at risk because of a lack of holistic archiving,
and c) source code is being cited more in our scholarly literature. To understand the scope of source code citations and quantify the risk of loss, we analyzed a corpus of scholarly publications and the URIs to GHPs that the publications contain.

\section{Methodology}
We decided to analyze 
the arXiv and PubMed Central corpora as a representative sample of scholarly publications across Science, Technology, Engineering, and Math (STEM) disciplines, in order to understand how scholarly code is being referenced over time and, therefore, both woven into the fabric of our scholarly conversation and worthy of preservation. arXiv is one of the largest and most popular pre-print services, and the corpus contains over 2 million submissions \cite{fromme-cornell2022} from eight disciplines: physics, mathematics, computer science, quantitative biology, quantitative finance, statistics, electrical engineering and systems science, and economics. The arXiv corpus does not allow for anonymous submissions, is publicly available, and is accessible for programmatic acquisition and analysis. The PubMed Central (PMC) corpus contains publicly available full-text articles from a wide range of biomedical and life sciences journals. Only peer-reviewed journals are eligible for inclusion.\footnote{https://www.ncbi.nlm.nih.gov/pmc/pub/addjournal/} The most prevalent journals in the corpus are listed in Table \ref{tab:pmc_corpus} along with the number of articles in the corpus, the date of the first article available, and the date of the latest article. The size and availability of the arXiv and PMC corpora make them suitable for the purposes of our study.  

\begin{table}
  \centering
  \begin{tabular}{|l|r|r|r|}
    \hline
    Journal & Articles & Earliest & Latest\\
    \hline
    The Indian Medical Gazelle & 29,143 & 1866 & 1955\\
    The Journal of Cell Biology & 24,349 & 1962 & 2022\\
    The Journal of Experimental Medicine & 24,207 & 1896 & 2022\\
    BMJ Open & 21,565 & 2011 & 2022 \\
    Edinburg Medical Journal & 20,160 & 1855 & 1954\\
  \hline
\end{tabular}
\caption{Five most popular journals in the PMC corpus}
\label{tab:pmc_corpus}
\end{table}

In April 2007, the arXiv identifier scheme changed to accommodate a larger number of submissions and to address other categorization issues.\footnote{https://arxiv.org/help/arxiv\_identifier} We decided that beginning our arXiv corpus in April 2007 would suit our analysis, because three of the four repository platforms that we analyzed began after 2007. Each pre-print in arXiv can have multiple versions. When an author uploads a new version of the pre-print to the service, the version number increments by one. All versions of a pre-print are accessible in arXiv via a version-specific URI. For our analysis, we considered only the latest version of each submission, assuming that the final submission was the most complete and most representative of the author's intentions. With only the latest version of each submission, our arXiv corpus contained 1.56 million publications in PDF format from April 2007 to December 2021. 

The PMC corpus includes articles from the late 1700s to present. In order to more easily compare the corpora and because, as previously noted, three of the four repository platforms we analyzed began after 2007, we decided that beginning our PMC corpus in January 2007 was appropriate for our analysis. Additionally, the PMC corpus  separates articles that are available for commercial use from those that are only available for non-commercial use. We chose to analyze the articles that were only available for non-commercial use. Our PMC corpus contained 1.08 million publications in PDF format from 2007 to 2021. Between the arXiv and PMC corpora, we analyzed 2,641,041 publications. 

A study by Milliken \cite{iasge_enviro_scan} conducted initial testing of GitHub, GitLab, SourceForge, and Bitbucket to understand the archival quality available through Brozzler (Archive-It's crawler), a Standard crawler (Heritrix and Umbra), and Memento Tracer. Our project is a continuation of that study and, as a result, we chose to analyze the use of those four GHPs in the arXiv and PMC corpora. The GHPs are summarized in Table \ref{tab:repos}.

\begin{table}
  \centering
  \begin{tabular}{|l|r|l|l|}
    \hline
    Name & Start Date & Protocol & URI\\
    \hline
    SourceForge & 1999 & git and SVN & \url{https://sourceforge.net}\\
    Bitbucket & 2008 & git & \url{https://bitbucket.org}\\
    GitHub & 2008 & git & \url{https://github.com}\\
    GitLab & 2014 & git & \url{https://gitlab.com}\\
  \hline
\end{tabular}
\caption{Repository Platforms}
\label{tab:repos}
\end{table}

URIs are not exclusively found in the References section of a publication; they also commonly appear in footnotes and the body of the text. To extract all of the URIs in each publication, regardless of location, we leveraged two Python libraries: PyPDF2\footnote{https://pypi.org/project/PyPDF2/} and PyPDFium2.\footnote{https://pypi.org/project/pypdfium2/} We used PyPDF2 to extract annotated URIs and PyPDFium2 to extract URIs from the PDF text. We followed a similar URI characterization method as that done by Klein et al. \cite{klein-plos2014} who identified URIs to ``Web at large" resources in-scope for their study. Since we are investigating links to GHPs, our primary goal with extraction was to identify URIs to one of the four GHPs. However, we also identified URIs to the Web at large to provide context for the frequency and use of URIs to the GHPs. To do this, we filtered out a number of URIs that were out of scope for this study. We dismissed URIs with a scheme other than HTTP or HTTPS, including localhost and private/protected IP ranges. We also dismissed URIs to arXiv, Elsevier RefHub,\footnote{https://refhub.elsevier.com} CrossRef Crossmark \cite{hendricks-crossref-2020}, and HTTP DOIs and, as such, follow the definition of URIs to ``Web at large" resources that are in-scope for our work. DOIs resolve to artifacts, most commonly papers but increasingly also to data (e.g., via Dryad) and source code (e.g., via Zenodo). Links to Elsevier RefHub and CrossRef Crossmark function similarly to DOIs and are often added by the publisher. We decided to exclude DOI and DOI-like references following Klein et al.'s assumption that, for the most part, such artifacts are in-scope for existing archiving and preservation efforts such as LOCKSS \cite{reich-dlib2001}, CLOCKSS \cite{reich-serials2008}, and Portico \cite{fenton-serials2006}. Our source code is available on GitHub \cite{Extract-URLs}.

After extracting URIs from the PDFs in our corpora, we found 7,746,682 in-scope URIs: 4,039,772 URIs from the arXiv corpus and 3,706,910 URIs from the PMC corpus. Out of 2.64 million files, 1,439,177 files ($54.06\%$) contained a URI. Once we had collected all of the URIs from the PDFs, we used regular expressions to filter and categorize the URIs that referenced one of the four GHPs. As a result, URIs to repository pages with custom domain names \cite{custom-domain} were not captured. We found a total of 253,590 URIs to one of the four GHPs: 231,206 URIs from the arXiv corpus and 22,384 URIs from the PMC corpus. Additionally, we found that 92.56\% of the GHP URIs were still available on the live Web. All GHP URIs in a publication have been deemed by the authors to be important enough for inclusion in the publication. As a result, we do not differentiate links to GHPs regardless of link depth or location in the publication. Inclusion of a GHP URI does not indicate an authorship or ownership claim. GHP URIs in a publication indicate that a resource either 1) impacted the work presented in the publications or 2) was a product of the study. Both cases communicate the importance of the repository and need for preservation. The number of URIs for each GHP are shown in Table \ref{tab:repo_count}. The URIs to GitHub account for 92.3\% of the URIs to one of the four GHPs. 

\begin{table}
    \centering
    \begin{tabular}{|l|r|r|}
    \hline
    Repository Platform & arXiv & PMC\\
    \hline
    GitHub & 215,621 & 18,471\\
    SourceForge & 9,412 & 3,309\\
    Bitbucket & 3,525 & 437\\
    GitLab & 2,648 & 167\\
  \hline
    \end{tabular}
    \caption{Number of references to each GHP in the arXiv and PMC corpora}
    \label{tab:repo_count}
\end{table}

\section{Results}
\begin{figure}
    \centering
    \includegraphics[width=\linewidth]{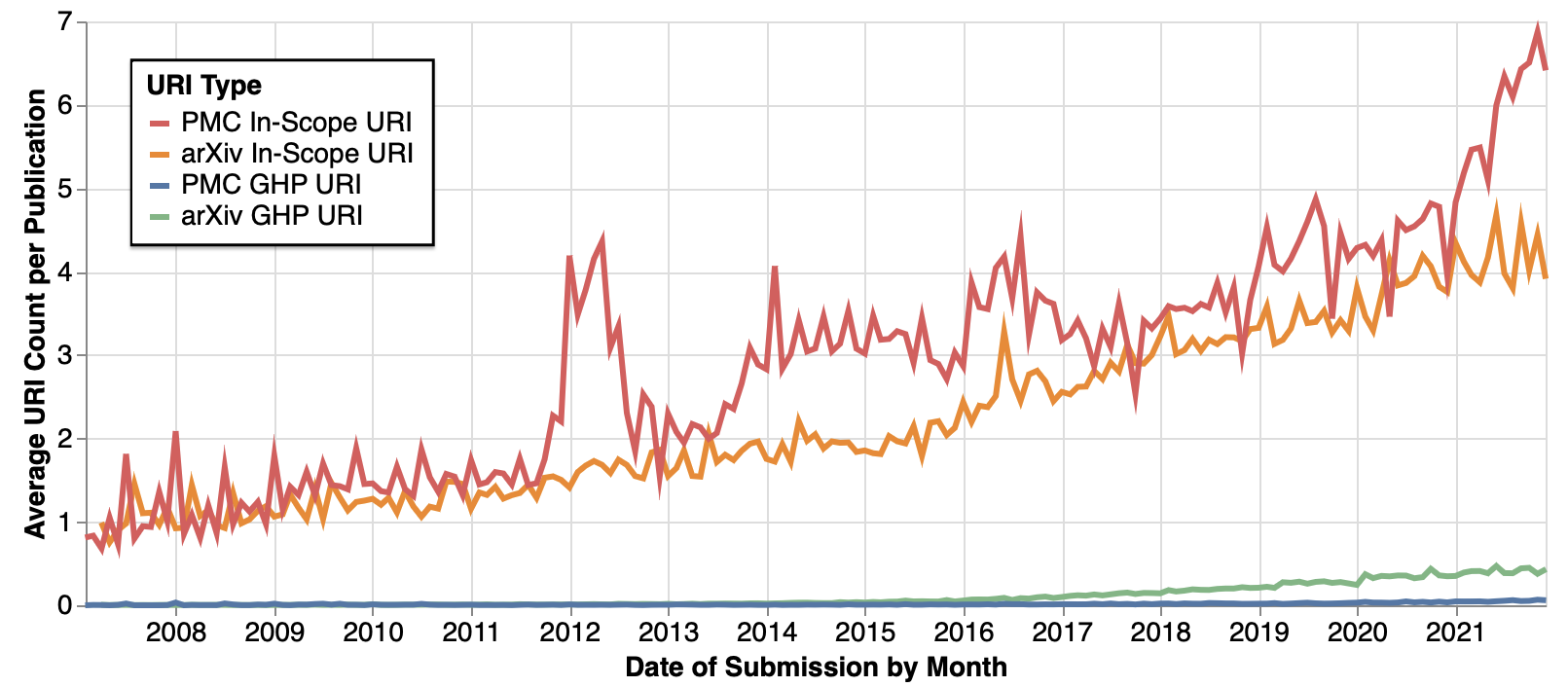}
    \caption{The average number of in-scope URIs and URIs to repository platforms per publication over time}
    \label{fig:combo_urls}
\end{figure}

By extracting URIs for the four repository platforms, we made a number of interesting observations. As shown in Figure \ref{fig:combo_urls}, we found a continuation of the significant increase in the prevalence of URIs in publications that Klein et al. \cite{klein-plos2014} found in 2014. Figure \ref{fig:combo_urls} shows the average number of in-scope URIs and the average number of URIs to one of the four GHPs in each publication by month of submission for both the arXiv and PMC corpora. The URIs to one of the four GHPs are a subset of in-scope URIs extracted from the publications. From 2007 to 2021, the average number of URIs per publications has steadily risen. In 2007, publications contained an average of 1.02 URIs. In 2021, publications contained an average of 5.06 URIs. The average number of in-scope URIs in each publication is indicated by the red and orange lines in Figure \ref{fig:combo_urls}. 

\begin{figure}[h]
\centering
\begin{subfigure}{\textwidth}
    \includegraphics[width=\textwidth]{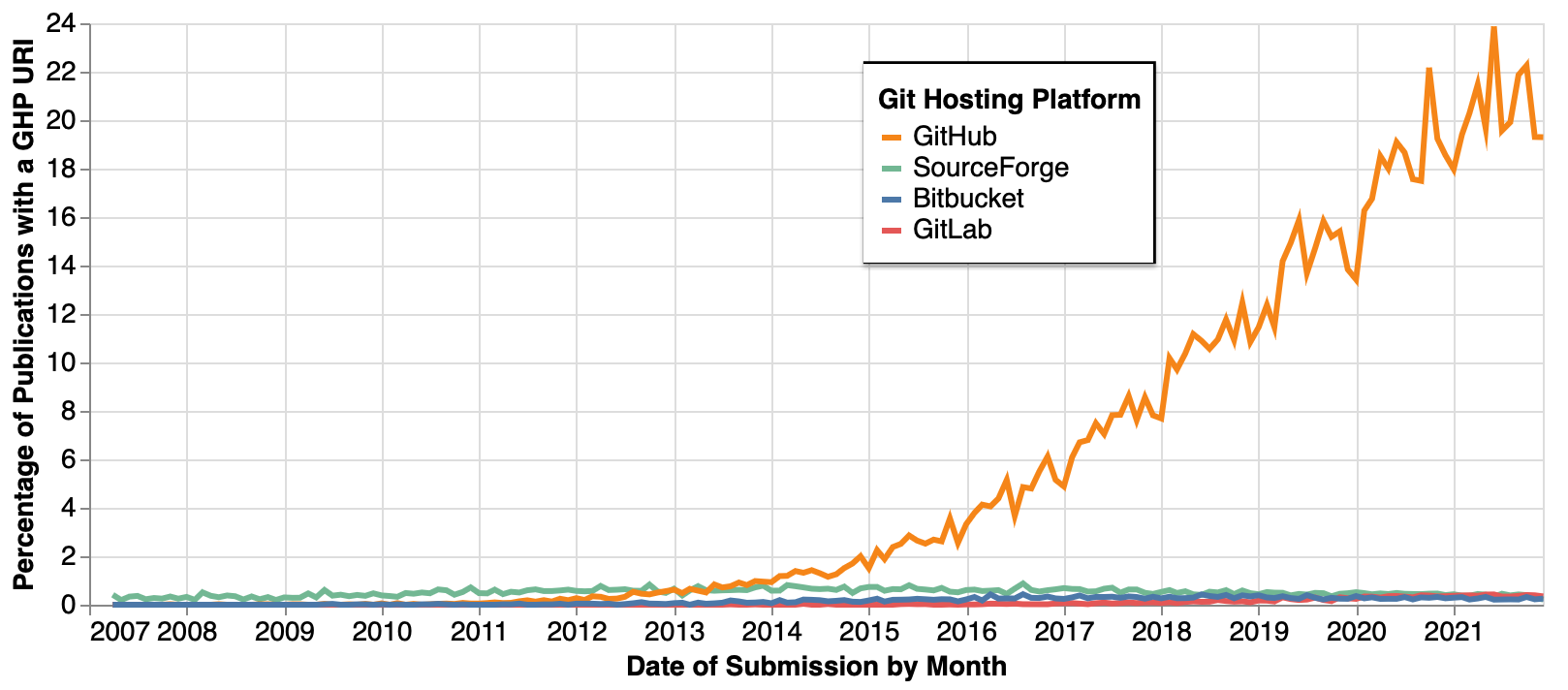}
    \caption{arXiv corpus}
    \label{fig:arxiv_repo_urls}
\end{subfigure}
\hfill
\begin{subfigure}{\textwidth}
    \includegraphics[width=\textwidth]{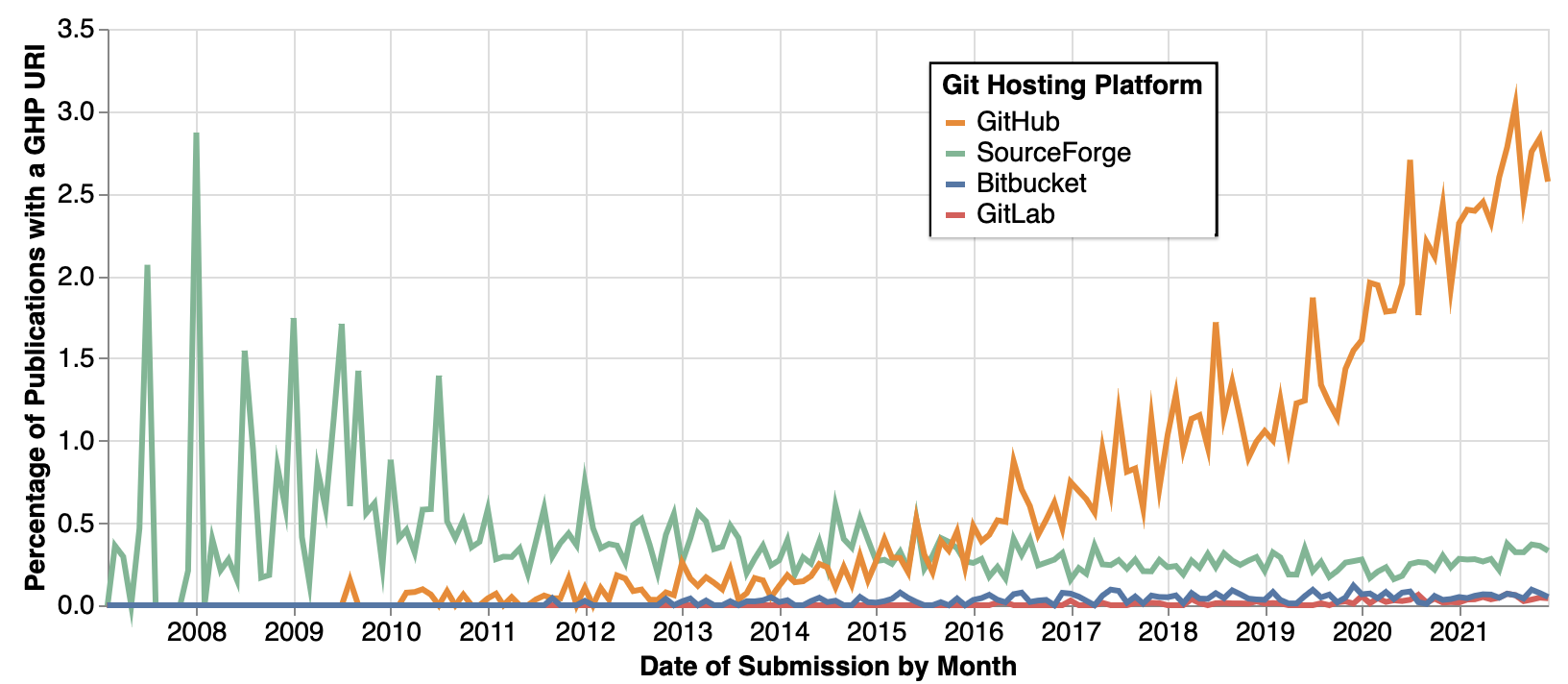}
    \caption{PMC corpus}
    \label{fig:pmc_repo_urls}
\end{subfigure}
        
\caption{The percentage of publications with a URI to a repository platform over time. Please note that the graphs are not on the same y-axis scale.}
\label{fig:repo_urls}
\end{figure}

While the prevalence of URIs in general has increased, the number of URIs to repository platforms has also grown from 2007 to 2021. Just as there was a shift from not including Web resources in scholarly publications to including Web resources, there has also been a shift to referencing repository platforms in scholarly publications. Figure \ref{fig:repo_urls} shows that references to GitHub have steadily risen from 2014 to 2021 while the frequency of references to the other three platforms have remained low during that time period. In the arXiv corpus shown in Figure \ref{fig:arxiv_repo_urls}, less than 1\% of publications contain a URI to GitLab, Bitbucket, or SourceForge in any given year from 2007 to 2021. However, an average of 20\% of publications contained a URI to GitHub in 2021. The PMC corpus in Figure \ref{fig:pmc_repo_urls} shows an initial prevalence of SourceForge beginning in 2007, but it is replaced by GitHub in 2015. Both graphs show a steady increase in the use of GitHub URIs in scholarly publications. Like URIs to the Web at large, URIs to repositories contribute to the context and argument of the publication. As the prevalence of GitHub URIs in publications increases, so does the importance of archiving source code repositories with its scholarly ephemera.

\begin{figure}[htbp]
\centering
\begin{subfigure}{\textwidth}
    \includegraphics[width=\textwidth]{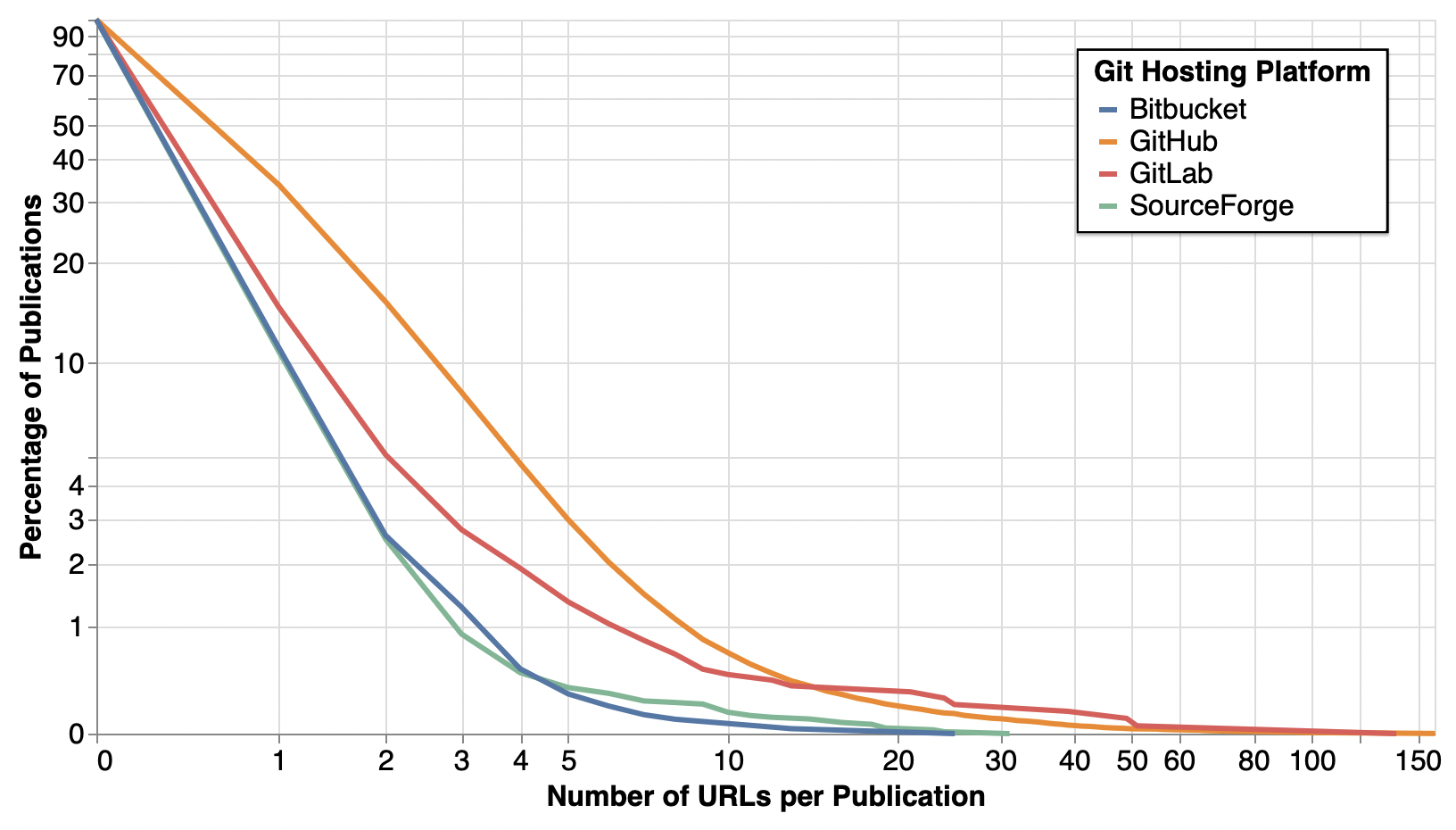}
    \caption{arXiv corpus}
    \label{fig:arxiv_ccdf}
\end{subfigure}
\hfill
\begin{subfigure}{\textwidth}
    \includegraphics[width=\textwidth]{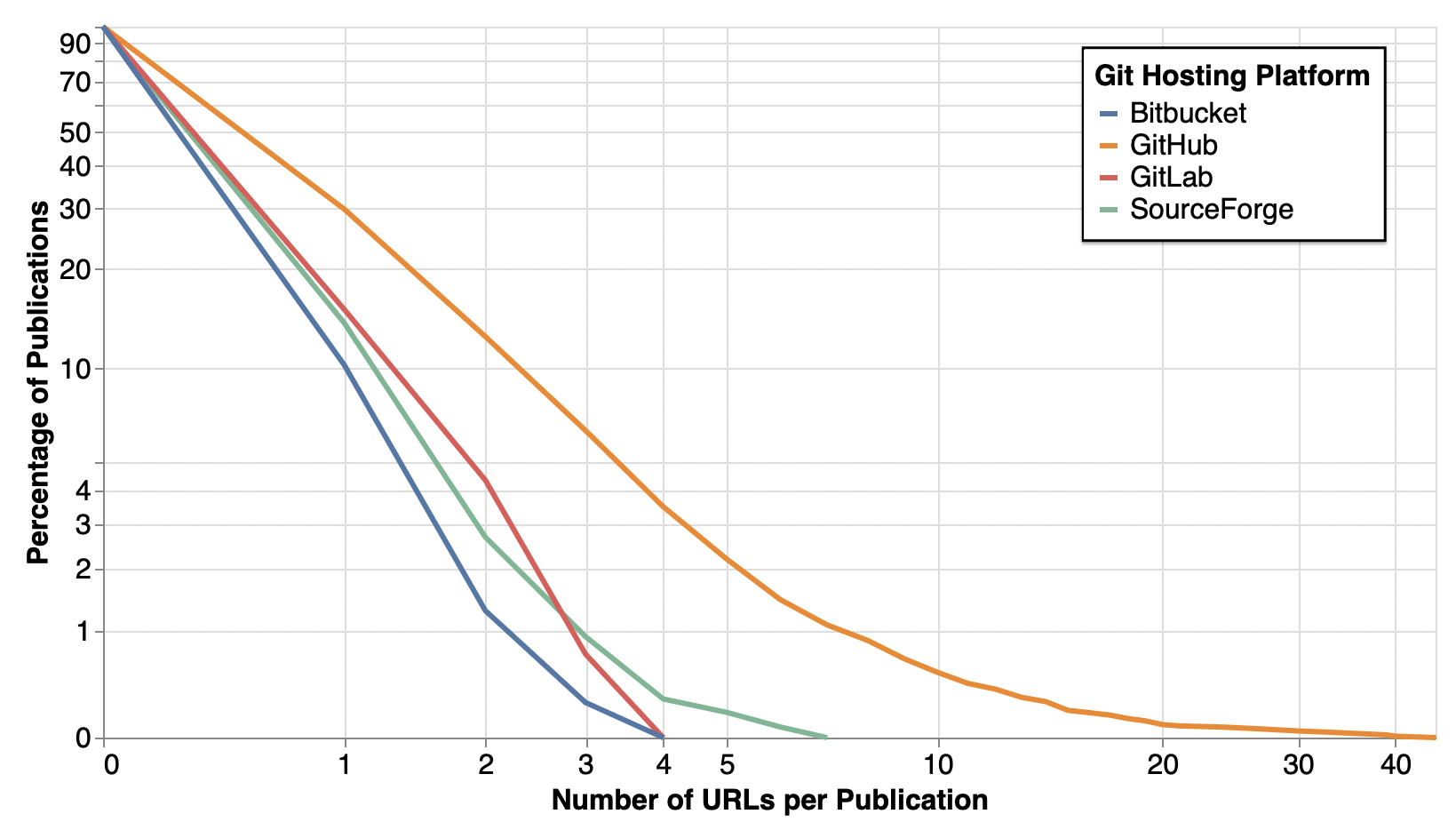}
    \caption{PMC corpus}
    \label{fig:pmc_ccdf}
\end{subfigure}
        
\caption{If a publication links to GHP, how many links does it have? This figure is a Complementary Cumulative Distribution Function (CCDF) graphing the frequency of GHP URIs in publications with 1 or more GHP URI}
\label{fig:ccdf}
\end{figure}

Additionally, while 67\% of publications only reference a given repository once, 45,780 publications reference a given platform's holding more than once. Figure \ref{fig:ccdf} shows the frequency of GHP URIs in publications that contain one or more GHP URI. For example, as shown in Figure \ref{fig:arxiv_ccdf}, of the 125,711 publications in the arXiv corpus that reference GitHub, 83,328 publications ($66.3\%$) reference GitHub once, 42,383 publications ($33.7\%$) reference GitHub more than once, and 863 publications ($0.687\%$) reference GitHub more than ten times. We manually inspected a sample of the publications with the most URIs to one of the four GHPs and found these publications tend to detail a software product or provide an overview of a topic, such as survey paper. The top three publications containing the most URIs to a GHP include 153 \cite{dhole-arxiv2021}, 160 \cite{agol-arxiv2021}, and 896 \cite{truyen-arxiv2021} URIs to GitHub. Dhole et al. \cite{dhole-arxiv2021} developed a software product and included URIs to the implementation of the features listed in the publication. Agol et al. \cite{agol-arxiv2021} created an open-source package and linked to the implementation of the algorithms and processes described in the publication. Truyen et al. \cite{truyen-arxiv2021} wrote a survey paper comparing frameworks. A majority of the frameworks surveyed are documented in GitHub, so the survey contains numerous URIs to the documentation. The publication by Truyen et al. with 896 URIs to GitHub is not included in Figure \ref{fig:ccdf}, because it represents such a large outlier compared to the other publications in the corpus.

As shown in Figure \ref{fig:pmc_ccdf}, of the 11,386 publications in the PMC corpus that reference GitHub, 7,983 publications ($70.1\%$) reference GitHub once, but 3,403 publications ($29.9\%$) reference GitHub more than once and 60 publications ($0.527\%$) reference GitHub more than ten times. The top four publications with the most URIs to a GHPs contain 39 \cite{kuo-pmc2019,kayani-pmc2021}, 40 \cite{chen-pmc2021}, and 45 \cite{yang-pmc2021} URIs to GitHub. Like the arXiv corpus, each of these four publications provides a survey of the computation tools available in a given discipline.

Publications with multiple references to GitHub imply the repositories have significant value and relevance for the authors, indicating that they deemed the repository contents important to the content of the publication. As a result, these repositories should be preserved in archives to guarantee that future readers can access the publication's full context.

\begin{figure}
    \centering
    \includegraphics[width=\textwidth]{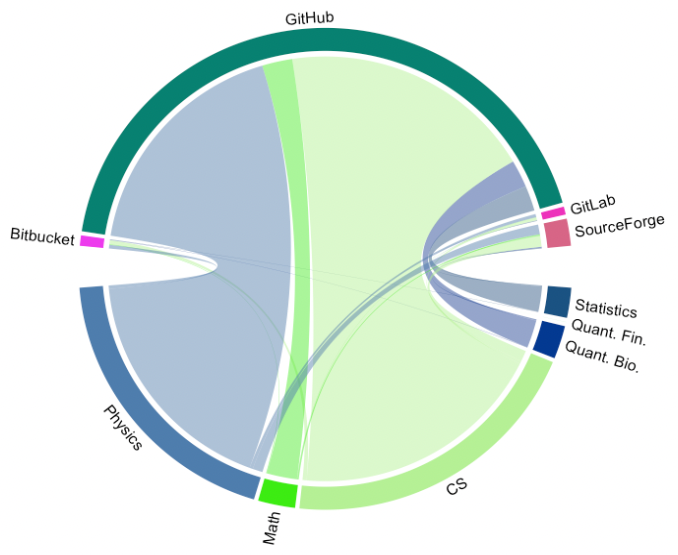}
    \caption{Mapping the number of links to a GHP (top half of the diagram) by discipline (bottom half) for the arXiv corpus}
    \label{fig:chord_diagram}
\end{figure}

We also analyzed the use of URIs to GHPs by discipline for the arXiv corpus. When submitting an article to arXiv, authors are prompted to select the primary discipline of the article. We used the metadata associated with each article to map each discipline to the four GHPs based on the number of URIs to each GHP. Figure \ref{fig:chord_diagram} shows a visualization of the relationship between GHPs and STEM disciplines. Computer Science and Physics contain the highest number of URIs to a GHP. Considering the prevalence of software products and models in the Computer Science and Physics disciplines, these results are not surprising.

\section{Discussion}
We analyzed the holdings of the arXiv and PMC corpora, but other corpora that service a wider variety of disciplines could provide additional perspectives. Additionally, authors must submit their paper to the arXiv corpus. This could create another source of bias in that authors must be able to navigate the submission process and must choose to submit their publication. Authors who intentionally submit their paper to arXiv are proving that they value open source and resource sharing, so this may be one reason that links to GHPs are more prevalent in the arXiv corpus. The PMC corpus is an example of a corpus that does not require action by the authors. Journals apply to be included in the PMC archive and all articles from the journal are automatically included. In future work, we will look at aggregating additional corpora to obtain a more representative sample of disciplines.

This analysis can also be used to supplement software preservation efforts. Curators and archivists could use these extraction methods to identify potential software of interest for their collections. Using different methods, these URIs can then be used to seed the archiving process. For instance, these URIs could be used with the Memento Tracer framework\footnote{http://tracer.mementoweb.org} proposed by Klein et al. \cite{klein-dlok2019}, which aims to strike a balance between scalability and quality for archiving scholarly code with its scholarly ephemera at scale. Memento Tracer allows users to create a heuristic called a trace, which can be used for for a class of Web publications. In testing the Memento Tracer framework, Klein et al. \cite{klein-dlok2019} was able to capture 100\% of the expected URIs for 92.83\% of the GitHub repositories in a given dataset. Additionally, the Memento Tracer framework was only 10.17 times slower than a typical crawler, while a comparable solution by Brunelle et al. \cite{brunelle-jcdl2017} was 38.9 times slower than a typical crawler. In future work, URIs derived using methods proposed in this study could be used to test the effectiveness of different archiving approaches at scale.

\section{Conclusions}
Sharing scholarly source code publicly is helpful for reproducing and verifying others' work, understanding the history of science, and facilitating far-reaching collaborations. The increase in scholars' usage of both VCS and GHP will aid open science. However, reference rot plagues the live Web, making archival efforts increasingly important. Citations to code in scholarly work serve as signals that these resources must be archived to preserve the scholarly record.  

For this study, we used the arXiv and PMC corpora to analyze the use of URIs to the Web at large and to Bitbucket, SourceForge, GitLab, and GitHub in scholarly publications. Our research found that scholarly publications increasingly reference the Web and software repository platforms. On average, a publication contains five URIs. Additionally, one out of five publications contains a reference to GitHub in the arXiv corpus. Each reference to a GHP's URI illustrates that content on these platforms constitutes an essential part of the context of the scholarly publication, and highlights the need to archive source code and accompanying scholarly ephemera hosted on GHPs.

\bibliographystyle{splncs04}
\bibliography{paper}
\end{document}